# Volumetric $B_1^+$ field homogenization in 7 Tesla brain MRI using metasurface scattering

Gyoungsub Yoon[1], Sunkyu Yu[2†], Jongho Lee[3], and Namkyoo Park[1*]

[1]Photonic Systems Laboratory, Department of Electrical and Computer Engineering, Seoul National University, Seoul 08826, Korea

[2]Intelligent Wave Systems Laboratory, Department of Electrical and Computer Engineering, Seoul National University, Seoul 08826, Korea

[3]Laboratory for Imaging Science and Technology, Department of Electrical and Computer Engineering, Seoul National University, Seoul, Korea

E-mail address for correspondence: [†]sunkyu.yu@snu.ac.kr, [*]nkpark@snu.ac.kr

**Abstract**

Ultrahigh field magnetic resonance imaging (UHF MRI) has become an indispensable tool for human brain imaging, offering excellent diagnostic accuracy while avoiding the risks associated with invasive modalities. When the radiofrequency magnetic field of the UHF MRI encounters the multifaceted complexity of the brain—characterized by wavelength-scale, dissipative, and random heterogeneous materials—detrimental mesoscopic challenges such as $B_1^+$ field inhomogeneity and local heating arise. Here we develop the metasurface design inspired by scattering theory to achieve the volumetric field homogeneity in the UHF MRI. The method focuses on finding the scattering ansatz systematically and incorporates a pruning technique to achieve the minimum number of participating modes, which guarantees stable practical implementation. Using full-wave analysis of realistic human brain models under a 7 Tesla MRI, we demonstrate more than a twofold improvement in field homogeneity and suppressed local heating, achieving better performance than even the commercial 3 Tesla MRI. The result shows a noninvasive generalization of constant intensity waves in optics, offering a universal methodology applicable to higher Tesla MRI.

# Introduction

Magnetic resonance imaging (MRI) is one of the most widely used neuroimaging modalities for noninvasive and precise measures of the central nervous system. The major determinant for advanced MRI is the main static magnetic field $B_0$, which is a key parameter of the signal-to-noise ratio (SNR)[1,2] and spatial resolution[3,4]. By increasing $B_0$ over 7 Tesla (T), the ultrahigh field (UHF) MRI enables the clinical diagnostics of intractable brain diseases—Alzheimer[5] and Parkinson[6] diseases—and in the preventive screening and global brain classification[7].

A new challenge arising in the UHF MRI is to overcome the issues stemming from the increased Larmor frequency of the transmit radiofrequency (RF) magnetic field $\mathbf{B}_1^+$, which is proportional to the static magnetic field $B_0$. In the $B_0 \geq 7\text{T}$ brain MRI, the wavelength of the $\mathbf{B}_1^+$ field inside the human tissue is comparable to or smaller than a human head, leading to the emergence of electromagnetic multimodes and their complex interferences, and thus, field inhomogeneity inside this biological resonator[8-10]. Because $\mathbf{B}_1^+$ field homogenization is essential for high SNR and uniform imaging contrast[9,11,12], advanced UHF MRI requires addressing the following electromagnetic problem: achieving $\mathbf{B}_1^+$ field homogenization and low electromagnetic energy absorption inside a highly deformed, wavelength-scale, and multimode resonator, which is filled with random heterogeneous and dissipative materials and is excited by the stationary outer RF excitation coil. This intricate problem furthermore pertains to mesoscopic systems characterized by subwavelength inhomogeneity, wavelength-scale footprints, and nonergodicity, which hinder the exclusive application of traditional theories in optics, such as the effective medium theory or multiple scattering theory.

Achieving field homogenization is not the only interest in the UHF MRI application, as evidenced by efforts to realize constant-intensity waves in gain and loss materials[13-15] and nonlinear chaotic resonators[16]. However, due to the restrictions of noninvasive measurement, the delicacy of biological tissues, and the immovable RF-coil boundary conditions, almost all MRI field homogenization techniques resort to empirical or heuristic approaches. First, the active parallel RF transmission (pTx)[17,18] approach modulates RF current sources through multi-channel Tx coils to manipulate the $\mathbf{B}_1^+$ field distribution in the region of interest (ROI). This method is analogous to microwave phased-array antennas[19] or optical spatial light modulation[20,21], fundamentally originating from scattering theory with manipulations of multiple sources. Despite the advantages in flexibility, the pTx approach shares the same disadvantages as the related techniques[19-21], including implementation complexity due to the use of spatiotemporally varying RF fields, as well as an increase in the maximum electromagnetic absorption. Second, high permittivity materials (HPMs)[9,22] and metasurfaces[23,24] enable the tailoring of the field distribution under the principle of the effective medium theory. Although this passive approach allows for low-cost and geometrically simple access to field homogenization, the mesoscopic nature of the UHF MRI breaks the underlying assumption that treats the human brain as an object of low effective permittivity, preventing sufficient volumetric homogenization across this biological resonator.

Here, we develop the semi-analytical design procedure for this mesoscopic problem, achieving a twofold improvement in $\mathbf{B}_1^+$ homogeneity over the entire brain bulk with highly suppressed local heating in the 7T UHF MRI. The proposed method is based on finding the scattering ansatz, analogous to Green's function method, while simultaneously applying the pruning of impulse responses for stable implementation. The method leads to the systematic design of the metasurfaces that provide the volumetric tailoring of the wavefront in the entire ROI. The performance of our method is numerically demonstrated under the full-wave analysis for realistic and representative human brain models—the Duke model[25] for a male, the MIDA model[26] for a female, and the modified Duke modelling for a baby. Our proposal not only reveals the solution for the volumetric MRI homogenization which has remained a challenge for the past 20 years[12,27], but also provides a general recipe for the noninvasive field homogenization in optical mesoscopic environments.

## Results

### Design concept

We consider the traditional environment of the brain MRI operation, where a human head is surrounded by a multi-channel RF coil (Fig. 1a, left). The unit element of the coil is a transverse electromagnetic (TEM) resonator (Supplementary Note S1). The resonator is excited by a 50 Ω-resistive voltage source through an impedance matching circuit. The lumped elements of the matching circuit are tuned to suppress the return loss at the source termination to below –40 dB (Supplementary Note S2), guaranteeing maximum power transfer to the interior of the MRI. The 16-channel TEM coil, driven by sequentially phase-shifted 16 voltage sources, then collectively transfers most of the electromagnetic energy from the source to generate the circularly polarized magnetic field $\mathbf{B}_1^+$ for brain imaging (Supplementary Note S3).

To achieve the field homogenization of $\mathbf{B}_1^+$ in the UHF MRI, we employ the cylindroidal copper-wire metasurface inside the RF coil (Fig. 1a, right), which is designed to achieve the three-dimensional (3D) volumetric field homogenization of $\mathbf{B}_1^+$. Notably, the listed configuration includes spatial complexity in terms of wave physics. Although the coil and metasurfaces themselves approximately possess $z$-axis translational symmetry within a finite range and exhibit discrete rotational symmetry along the azimuthal axis, a human head corresponds to random heterogeneous materials with a number of material phases and the mixing of diverse morphologies (Fig. 1b). Therefore, devising the design strategy for the shape and location of the metasurface object is not a straightforward problem.

To resolve this spatially complex problem, we leverage a concept in scattering theory, evaluating the initial field configuration except for the metasurface and then calculating the induced scattering field from the inclusion of the metasurface. Considering metasurfaces as arbitrary scatterers at this point, we start from the Lippmann-Schwinger equation under highly simplified and analytically accessible conditions of isotropic and slowly-varying permittivity (Supplementary Note S4)

$$\mathbf{B}(\mathbf{r}) \sim \mathbf{B}_0(\mathbf{r}) + \frac{k_0^2}{4\pi} \int_\Omega \varepsilon_M(\mathbf{r}') \left( \mathbf{B}(\mathbf{r}') \odot \mathbf{G}(\mathbf{r} - \mathbf{r}') \right) d\mathbf{r}', \qquad (1)$$

where $\mathbf{B}_0(\mathbf{r})$ is the initial magnetic field induced by the RF coil, $\mathbf{B}(\mathbf{r})$ is the total magnetic field including the perturbation from the metasurface object, $\varepsilon_M(\mathbf{r})$ is the permittivity perturbation due to the metasurface object, $\mathbf{G}(\mathbf{r})$ is the Green's function for the impulse response with the Dirac delta function perturbation, $\Omega$ is the entire spatial domain inside the RF coil, and $\odot$ is the Hadamard product. The goal of designing metasurface objects is to achieve the distribution of $\varepsilon_M(\mathbf{r})$ to realize spatial homogeneity of $\mathbf{B}(\mathbf{r}) \sim \mathbf{B}_H$ with the minimum number of metasurface objects, where $\mathbf{B}_H$ is the spatially constant magnetic field. However, in addition to a well-known difficulty—recursive integral equation form—of the Born series, employing Eq. (1) directly to the optimization problem is intricate. It is because the Green's function $\mathbf{G}(\mathbf{r})$ cannot be obtained analytically due to the random heterogeneous profile of a human head and guaranteeing the convergence of the Born series in Eq. (1) is computationally expensive[28], especially for the free form and strong perturbation of $\varepsilon_M(\mathbf{r})$. Notably, the distribution of $\varepsilon_M(\mathbf{r})$ could be restrictive[29] when considering the practical implementation of noninvasive medical treatments.

As an alternative, we develop the design strategy applicable to general field homogenization problems (Fig. 1c), which employs the discretization, weighted magnetic ansatz, and pruning techniques that are inspired by Eq. (1). First, we assume the medically allowed subspace $\Omega_n$ ($n = 1, 2, …, N$), to which we will employ an $N$ number of metasurface objects. We then discretize $\varepsilon_M(\mathbf{r})$ to $\varepsilon_M(\mathbf{r} \in \Omega_n) = \varepsilon_n$ and $\varepsilon_M(\mathbf{r} \in [\cup_{n=1}^{N} \Omega_n]^c) = 0$. Second, under the philosophy of the Born approximation[30], we introduce the magnetic ansatz impulses $\mathbf{B}_n$ for each domain $\Omega_n$, as:

$$\mathbf{B}_n(\mathbf{r}) \sim \int_{\Omega_n} \varepsilon_n \left( \mathbf{B}_0(\mathbf{r}') \odot \mathbf{G}(\mathbf{r} - \mathbf{r}') \right) d\mathbf{r}'. \tag{2}$$

To resolve the absence of the well-defined Green's function $\mathbf{G}(\mathbf{r})$ for our problem, we substitute Eq. (2) with the numerical alternative, applying the full-wave analysis for each case of $\Omega_n$ with the fixed value of $\varepsilon_n$. To compensate for the Born approximation in Eq. (2), we employ the weight $c_n$ of each ansatz, as

$$\mathbf{B}(\mathbf{r}) \sim \sum_{n=1}^{N} c_n \mathbf{B}_n(\mathbf{r}) \tag{3}$$

where $c_n$ determines the contribution of each ansatz in the total field, including the modification of the incident field, as $\mathbf{B}_0(\mathbf{r}) \to c_0 \mathbf{B}_0(\mathbf{r})$. The field homogenization is achieved by minimizing the coefficient of variation (CV) defined for the target domain $\mathbf{V} \in \Omega$:

$$\text{CV} \triangleq \sqrt{\frac{1}{N_\mathbf{V}} \int_\mathbf{V} \left\| |\mathbf{B}(\mathbf{r})| - |\mathbf{B}_\mathrm{H}| \right\|^2 d\mathbf{r}}, \tag{4}$$

where $\mathbf{B}(\mathbf{r})$ is normalized and $N_\mathbf{V}$ is the volume of $\mathbf{V}$. Figure 1c (from c-1 to c-2) describes the two-dimensional (2D) toy model of this minimization process, applying the gradient descent method to obtain the weighting set $\{c_n\}$ for minimizing the CV.

Because the tractable form of the metasurface configuration is essential for practical implementation, we attempt to maintain an $M$ ($M < N$) number of metasurface objects. We therefore apply the "pruning" of the $N$ objects[31,32], by comparing the weight of each ansatz $c_n$. The pruning process corresponds to the characterization of the minimum number of eigenmodes, not only preventing overfitting in the optimization process but also enhancing the stability in the metasurface implementation.

After the pruning through the iteration process and remaining the $M$ objects at $\mathbf{\Omega}_m$ for the field homogeneity (Fig. 1c from c-2 to c-3), we choose to adjust $\varepsilon_m$ of $\mathbf{\Omega}_m$, to reflect the weight $c_m$ in the magnetic field scattering from the $m$th object. This optimization process inspired by scattering theory enables the systematic, efficient, and noninvasive design of the metasurface objects for the field homogenization, by only employing an $N$ number of full-wave simulations and practically valid setting of $M$ objects.

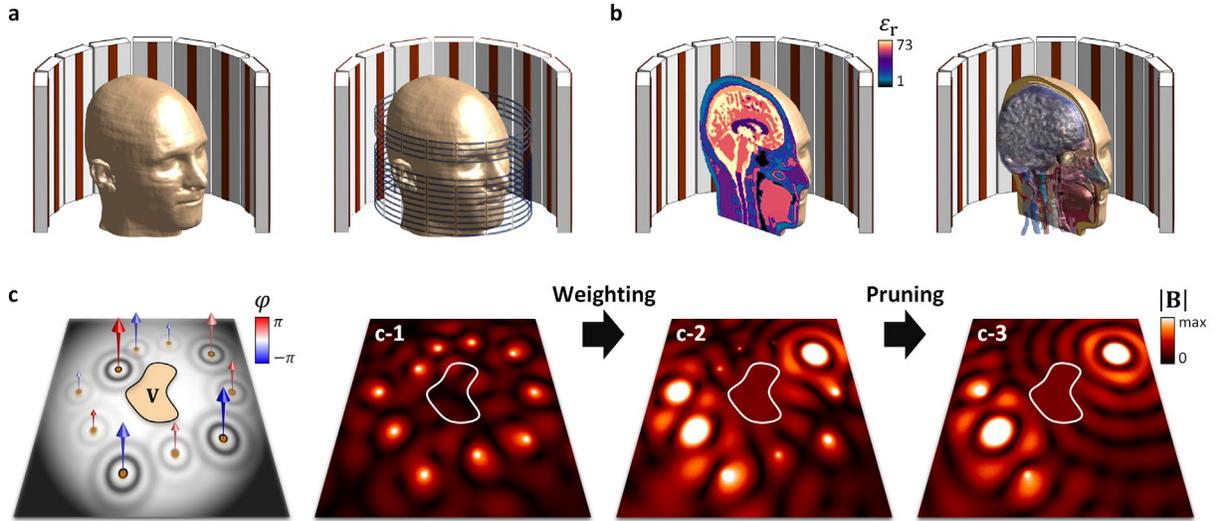

**Fig. 1. Metasurface UHF-MRI environment with optimized scattering. a,** A schematic of the MRI environment: the head model surrounded by the MRI coil without (left) and with (right) the metasurface objects. **b,** Random heterogeneous materials of a human head: cross-sectional profile of relative permittivity $\varepsilon_r$ at 300 MHz (left) and the blended morphologies of biological tissues (right). **c,** Illustration of the optimization process for the metasurface design. The arrow lengths and colors denote the amplitudes and phases of the weighted Dirac-delta-function-like point sources, which will be optimized to achieve the field homogenization inside the domain **V**. For the initial state (**c-1**), the optimization leads to different complex-valued weights to each point source (**c-2**). The pruning technique allows for removing relatively less important sources in the homogenization process (**c-3**), which enables more robust implementation. The schematics in **a** and **b** are obtained from the commercial software Sim4Life[33]. The result in **c** is obtained from the optimization process described in the next section.

**Optimization process**

We extend the toy model example depicted in Fig. 1c to the UHF MRI with copper-wire metasurface scattering. Instead of the ideal weighted Dirac-delta-function-like point sources in the toy model example, we employ passive impulse sources that model the scattering from each metasurface object. These magnetic ansatz impulses are located near a human head to couple with the reactive power inside the coil (Supplementary Note S5). The copper-wire metasurface objects are designed to achieve sufficiently high effective permittivity under curved geometry (Supplementary Notes S6-S8), which guarantees a broad range of weighting factors quantified by $\{c_n\}$ in Eq. (3).

Figure 2 describes the results of the optimization for the UHF MRI, targeting the minimization of $\mathrm{CV}(\{c_n\})$. Figure 2a shows the field distribution of the magnetic ansatz impulses **B**$_n$, which are achieved with the 3D finite-difference-time-domain (FDTD) simulations[33] each for $\varepsilon_\mathrm{M}(\mathbf{r}\in\mathbf{\Omega}_n) = \varepsilon_n$ and $\varepsilon_\mathrm{M}(\mathbf{r}\in[\cup_{n=1}^{N}\mathbf{\Omega}_n]^\mathrm{c})= 0$ (Supplementary Note S9 for the list of the target tissues). We adjust the vertical position of the metasurface object by 3 cm increments to define $\mathbf{\Omega}_n$, which controls the spatial distribution of the field concentration of each **B**$_n$.

Using the basis $\{\mathbf{B}_n\}$ and their complex-valued weights $\{c_n\}$, we attempt to achieve the homogenization inside an electromagnetically inhomogeneous brain object that composes the 3D volumetric ROI **V**. Figure 2b illustrates the process of the weighting and pruning, which exhibits the epoch-variation of the relative magnitudes of $\{c_n\}$. The result allows for the classification of $N$ metasurface objects according to their impacts on the homogenization of the magnetic field in the domain **V**: higher $|c_n|$ implying the importance of $\mathbf{B}_n$ in constructing $\mathbf{B}(\mathbf{r})$ close to $\mathbf{B}_H$. As shown in the below figures of the field distributions, the gradient-descent weighting process leads to the field homogenization for a given number of metasurface objects.

In the pruning process, we remain four objects based on CV($\{c_n\}$) in Fig. 2c. Notably, the values of the cost functions are maintained sufficiently small even when we apply the pruning technique, from $M$ = 6 to 4. The result demonstrates that our evolutionary optimization process provides the high performance metasurface objects with less complex geometry.

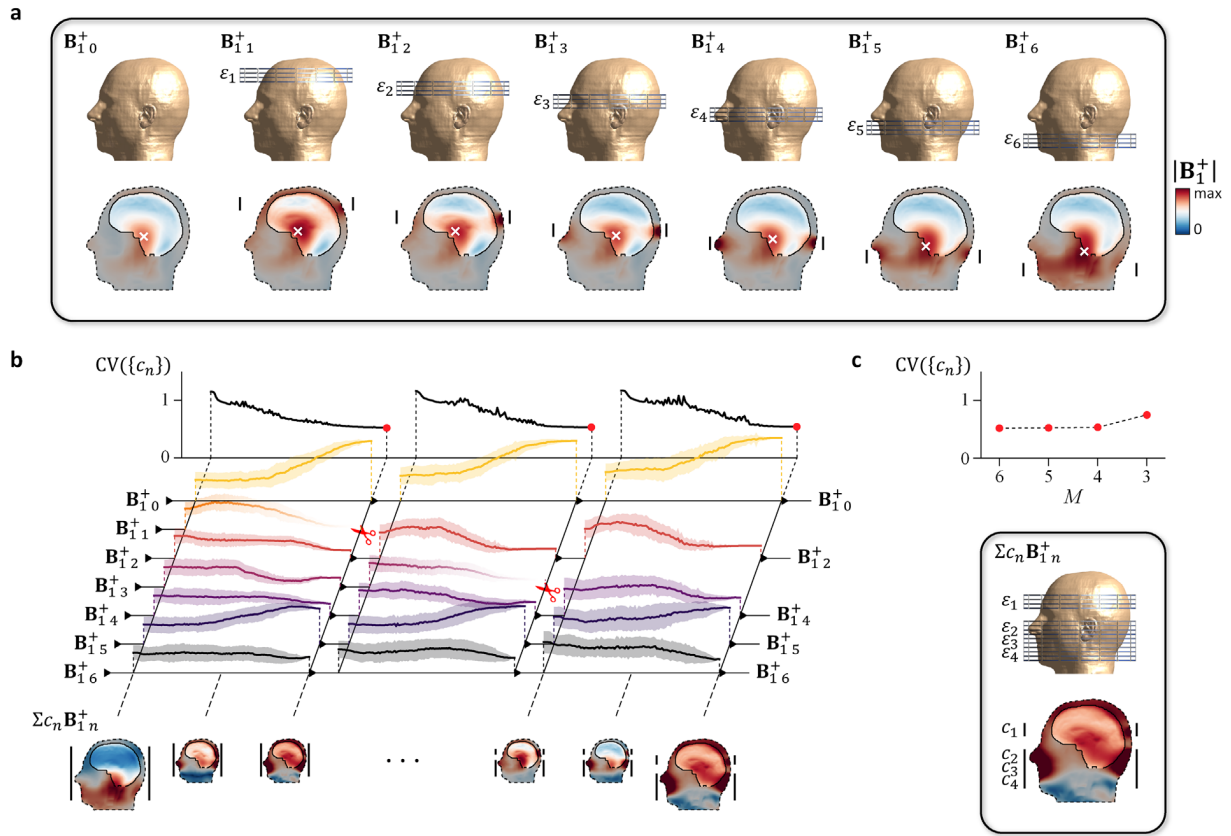

**Fig. 2. UHF MRI metasurface optimization. a,** Magnetic ansatz impulses $\{\mathbf{B}_1^+{}_n\}$ derived from the full-wave analysis[33] for the realistic head model (Duke model[25]): $\mathbf{B}_1^+{}_0$ induced solely by the RF coil, and $\mathbf{B}_1^+{}_n$ ($n$ = 1, 2, …, $N$) induced by both the RF coil and the $n$th-shifted metasurface object. Each object is shifted by every 3 cm along the vertical direction. The $n$th object possesses the effective permittivity of $\varepsilon_n$, which is the tunable parameter that controls $c_n$. In the field

distribution plot, the unshaded regions surrounded by the solid lines represent the target domain **V** for the field homogenization. **b,** The evolution of the normalized cost function CV($\{c_n\}$) according the optimization process defined by the gradient descent method and the pruning technique. The field distributions inside the brain (bottom) show the evolution of the field homogenization. Two red scissors illustrate the pruning process. Each line and width of $\{c_n\}$ denote the average and error bar obtained from an ensemble of 50 realizations of the uniformly random initial condition. **c,** The cost function against the pruning process and the resulting field distribution. $M$ ($M \leq N$) denotes the number of the metasurface objects after the pruning. The index of the metasurface object is reordered as $m = 1, 2, …, M$. In **a** and **c**, the white cross marker in each field profile indicates the point of the maximum field concentration within the ROI.

**Field homogenization**

The optimization process presented in Fig. 2 leads to the approximated profile of the $\mathbf{B}_1^+$ field for its homogenization using $M$ metasurface objects. However, the result itself does not directly render the complete design of metasurfaces due to our oversimplified Eq. (1) of the Born approximation, which neglects the recursive interactions between the scattering impulse responses from metasurface objects. As illustrated in Fig. 1b and the previous literature[23,24], the brain itself and the metasurfaces employed for $\mathbf{B}_1^+$ field homogenization usually have higher effective permittivities than the valid range of the Born approximation. In this context, the result in Fig. 2, which determines the pruned weighting factors $\{c_m\}$, represents only an approximate outline for designing metasurface objects following Supplementary Notes S6-S8.

Treating the solution of Fig. 2 as an ideal condition built upon scattering theory, we apply the depth-first search (DFS) algorithm to explore the optimum metasurface objects approaching the ideal condition. The algorithm involves manipulating the effective permittivity values $\{\varepsilon_m\}$ of the remaining metasurface objects (Fig. 3a, bottom), with the aim of achieving the optimal $\{c_m\}$ under the existence of the multiple scattering (Fig. 3a, top). Figure 3b shows a part of the similarity map between the normalized ideal field $\mathbf{B}(\mathbf{r};\{c_m\})$ and the normalized DFS-calculated field $\mathbf{B}(\mathbf{r};\{\varepsilon_m\})$ from the metasurface objects $\{\varepsilon_m\}$, where the similarity $\rho(\{\varepsilon_m\})$ is defined as:

$$\rho(\{\varepsilon_m\}) = 1 - \left| \int_V \mathbf{B}(\mathbf{r};\{c_m\}) \cdot \mathbf{B}^*(\mathbf{r};\{\varepsilon_m\}) d\mathbf{r} \right|. \tag{5}$$

When considering the match between the similarity map (Fig. 3b) and the CV map of $\mathbf{B}(\mathbf{r};\{\varepsilon_m\})$ (Fig. 3c) within the same parameter space, it is evident that the optimization process in Fig. 2 yields the well-established ansatz for field homogenization. The resulting 3D field homogenization is shown in Figs. 3d (para-sagittal planes), 3e (para-axial planes), and 3f (para-coronal planes). When compared with the results without metasurfaces (upper figures of Figs. 3d-f), the spatial localization of the field is evidently suppressed in all the observed cross-sections of a human head. Overall, we achieve a 40 % CV reduction for the Duke model, from 0.26 to 0.15, by employing metasurfaces.

Although our metasurface design is achieved with the optimization process for the Duke target model for males, the Duke-specified design is also valid for the head model of different scales: the MIDA model for females and the reduced Duke model for babies, achieving 46 % (from 0.37 to 0.20) and 44 % (from 0.31 to 0.17) CV reductions, respectively. Such superior adaptability demonstrates that our approach successfully achieves robustness to the perturbations in random heterogeneous materials of a human brain. This is one of the critical advantages of the hardware pruning technique[31] in addition to the simplicity of the system; the reduced number of impulse sources decreases the sensitivity of the designed optical interference pattern (Supplementary Note S10).

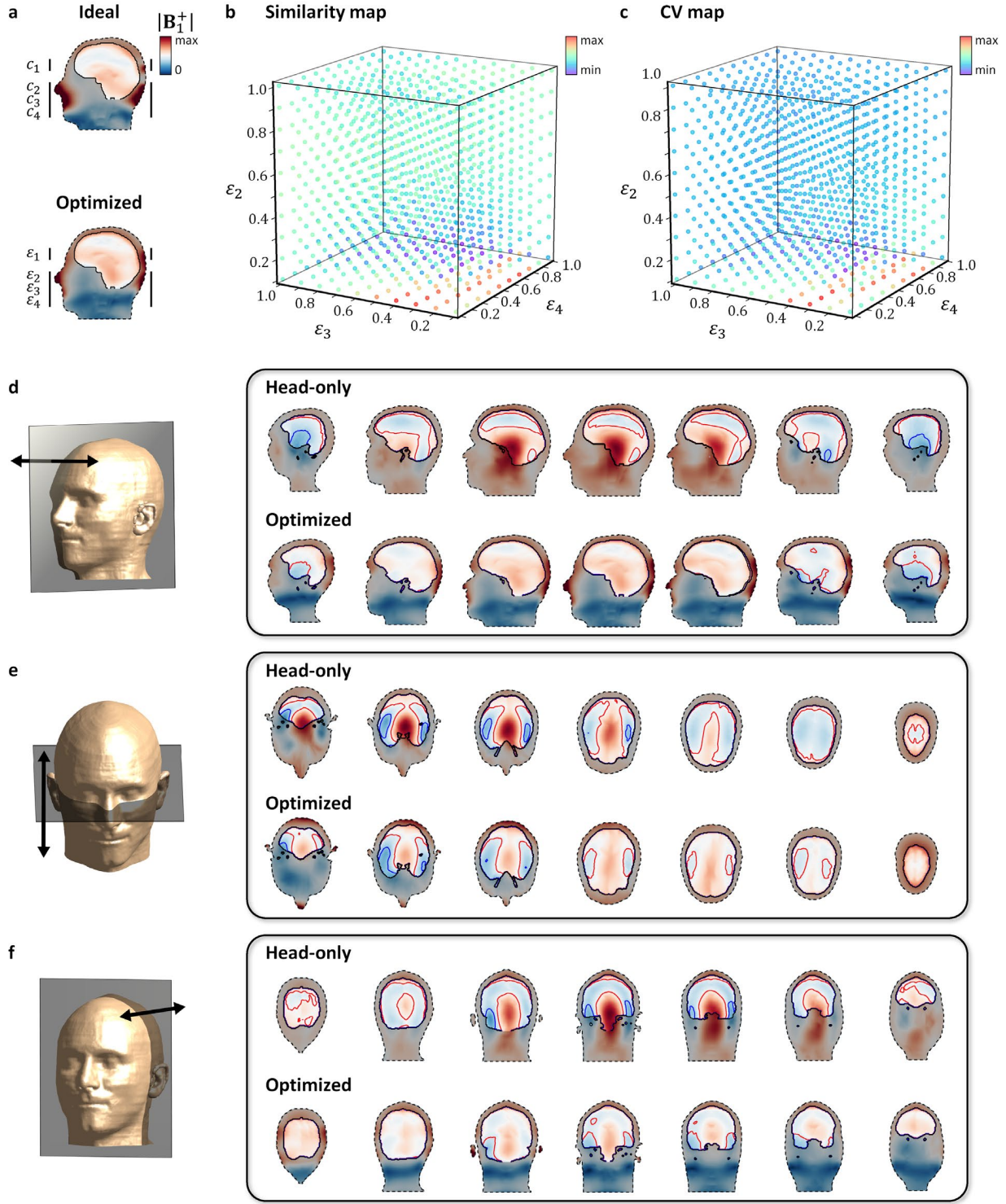

**Fig. 3. UHF MRI performance on field homogenization. a**, Midsagittal $\mathbf{B}_1^+$ cuts for comparing the ideal field $\mathbf{B}(\mathbf{r};\{c_m\})$ and the DFS-calculated field $\mathbf{B}(\mathbf{r};\{\varepsilon_m\})$. **b,c,** The similarity map $\rho(\{\varepsilon_m\})$ (**b**) and the CV map $\mathbf{B}(\mathbf{r};\{\varepsilon_m\})$ (**c**) for the 3D parameter space $(\varepsilon_2,\varepsilon_3,\varepsilon_4)$ at $\varepsilon_1 = 0.7$. **d-f,** Comparisons

of the $\mathbf{B}_1^+$ distributions in a human head without (upper) and with (lower) metasurface objects: Representative $\mathbf{B}_1^+$ slices at different para-sagittal planes (**d**), para-axial planes (**e**), and para-coronal planes (**f**). The $\mathbf{B}_1^+$ distributions of all cases are normalized to the volume integral of $|\mathbf{B}_1^+|$. The red and blue contours highlight the $|\mathbf{B}_1^+|/|\mathbf{B}_1^+|_{max} = 1/2$ and $1/3$, respectively.

**Suppressed heating**

For its clinically safe operations, the brain MRI puts stringent restrictions on the electromagnetic energy absorption in biological tissues to prevent a severe rise in temperature. This restriction is examined with another performance metric—specific absorption rate (SAR)[34]—which is defined locally as follows:

$$\text{SAR}(\mathbf{r}) = \frac{1}{M} \int_R \frac{\sigma(\mathbf{r}')}{2\rho(\mathbf{r}')} |\mathbf{E}(\mathbf{r}')|^2 \, dm(\mathbf{r}') \quad \text{[W/kg]} \tag{6}$$

Where $\sigma$ is conductivity (S/m), $\rho$ is the local mass density (kg/m$^3$), $dm(\mathbf{r})$ is the infinitesimal mass element at the position $\mathbf{r}$ (kg), $\mathbf{E}(\mathbf{r})$ is the induced electric field at $\mathbf{r}$, and $R$ is the spatial domain of 10 g mass, $M$, around $\mathbf{r}$.

Figure 4 shows the statistical analysis of the $|\mathbf{B}_1^+|$ field in the ROI and the SAR in the entire head for the UHF MRIs with and without the model-specified (Duke, MIDA, and reduced Duke) optimized metasurfaces (plane cut profiles in Supplementary Notes S11 and S12). We calculate the volumetric probability density functions for both the $|\mathbf{B}_1^+|$ and the local SAR for different head models. When compared to the case without metasurface objects (light green plots in Fig. 4a-f), the improvement both in the CV and SAR is evident for the Duke (Fig. 4a,d), MIDA (Fig. 4b,e), and the reduced Duke (Fig. 4c,f) models, achieving 48.53 % CV reduction, 22.98 % average SAR reduction, and 23.58 % peak SAR reduction on average compared to the head-only cases (Table 1 for details). Table 1 also shows that the achieved performance with the 7T MRI is almost close to that of the 3T MRI, overcoming the challenges originating from the more than two-fold increase of the TEM-coil operating frequency from 128 MHz to 300 MHz. We note that the Duke-specified metasurface is fairly compatible with the other head models also for the SAR performance (Supplementary Note S13), again demonstrating the robustness of our design strategy.

When considering the CV-targeting cost function in Eq. (4), the SAR suppression observed in Fig. 4d-f demonstrates a qualitative relationship between the homogenization of the magnetic field and the spectral characteristic of the consequent MRI system. Notably, the reduction in SAR, along with the flattened magnetic energy, originates from the separation of electric and magnetic energies (Supplementary Note S14). Because electric and magnetic energies are balanced in the on-resonance condition, our field homogenization design corresponds to a spectral-domain transition to an off-resonant state with a greater concentration to magnetic energy.

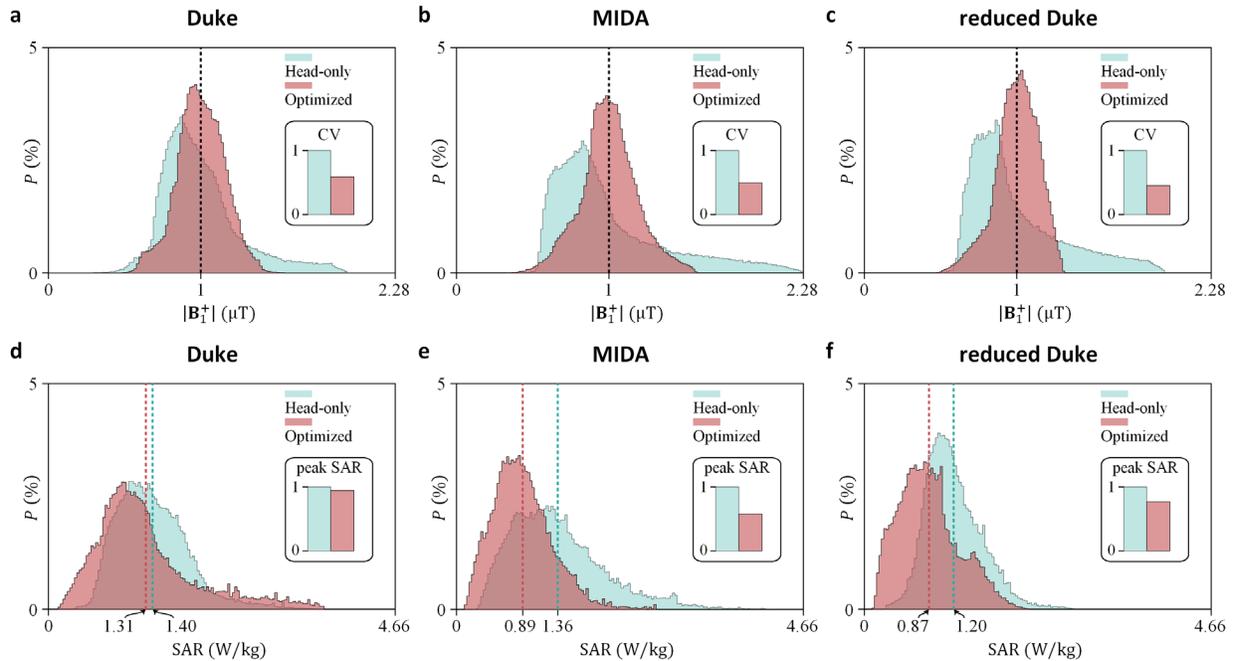

**Fig. 4. UHF MRI performance on global heat. a-c,** The volumetric probabilistic functions for the $|\mathbf{B}_1^+|$ with the average normalized to 1 µT, for the Duke (**a**), MIDA (**b**), and reduced Duke (**c**) models. **d-f,** The volumetric probabilistic functions for the SAR across the entire head of the Duke (**d**), MIDA (**e**), and reduced Duke (**f**) models. The improvement in the CV (**a-c**) and peak SAR (**d-**

**f**) achieved with the optimized metasurfaces are shown on each inset, where the head-only model is normalized. The dashed lines in **a-f** denote the measured averages.

**Table 1 | Field homogenization results for 7T and 3T MRIs.** CV and SAR values are compared for the Duke, MIDA, and reduced Duke models.

| Scheme | Model | CV | SAR [W/Kg] | |
| --- | --- | --- | --- | --- |
| | | | Average | Peak |
| 7T MRI without metasurfaces | Duke | 0.27 | 1.40 | 3.89 |
| | MIDA | 0.37 | 1.36 | 4.66 |
| | Reduced Duke | 0.31 | 1.20 | 2.85 |
| 7T MRI with metasurfaces | Duke | 0.15 | 1.31 | 3.69 |
| | MIDA | 0.18 | 0.89 | 2.67 |
| | Reduced Duke | 0.14 | 0.87 | 2.19 |
| 3T MRI without metasurfaces | Duke | 0.17 | 0.32 | 0.90 |
| | MIDA | 0.15 | 0.24 | 0.62 |
| | Reduced Duke | 0.11 | 0.20 | 0.47 |

In the volumetric field homogenization, achieving the stable performance against head postures is also critical for clinical stability during the medical treatment. To evaluate the robustness of our field homogenization scheme, we examine the UHF MRI operation under three different types of neck rotations: flexion, extension, and right lateral flexion, which are compared with no rotation ('neutral') case. For the rotation angle set to 5 degrees in each direction, the $B_1^+$ field homogenization is stably maintained inside the ROI under any type of rotation (Fig. 5a). In Fig. 5b-d, this stable performance with the optimized metasurfaces is confirmed with the improved CV values for different heal models: at least 38.5% (Duke), 49.2% (MIDA), and 53.2% (reduced Duke) reduction of CV for every rotation. This result demonstrates that our design strategy, which involves the optimized and simplified metasurfaces, enables stable UHF MRI operations in the dynamical environment of practical treatments.

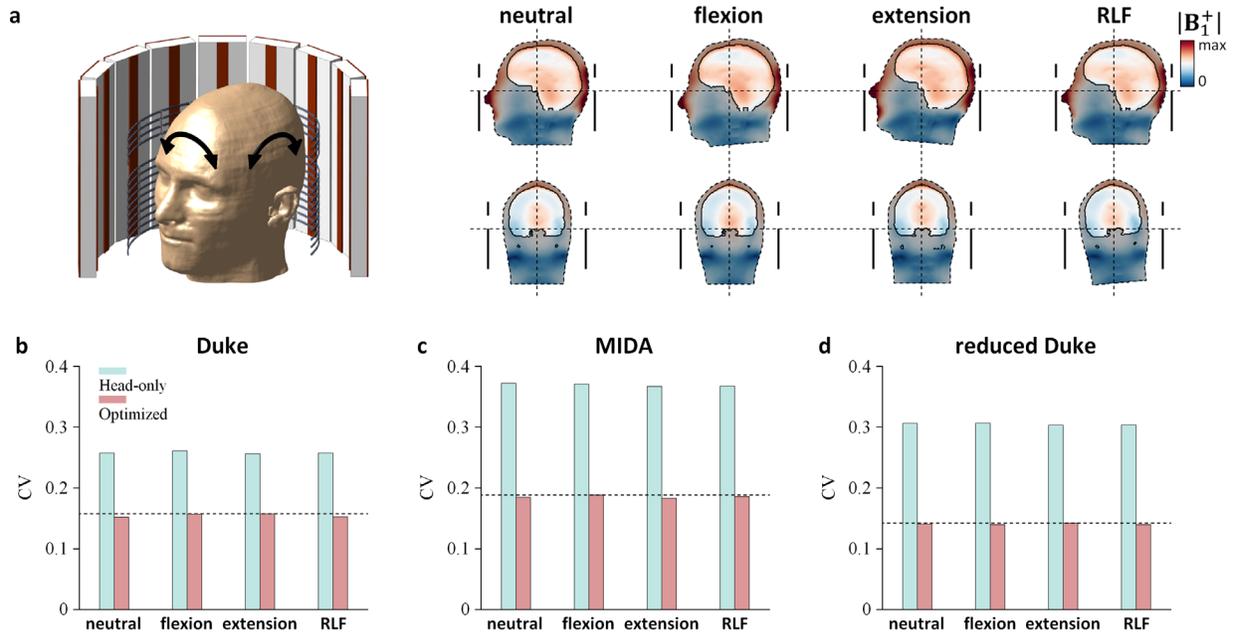

**Fig. 5. UHF MRI performance on field homogeneity against head rotations. a,** Robustness of $B_1^+$ homogenization with the optimized metasurfaces against three types of head rotations: front bending (neck 'flexion'), backward bending (neck 'extension'), and right lateral bending ('RLF') with respect to no rotation ('neutral'). The rotation angle is set to 5 degrees to the rotation axes (dashed lines). For each type of rotation, sagittal and coronal $B_1^+$ cuts are presented (**a**). **b-d,** CV for each rotation with ('Optimized') and without ('Head-only') the metasurfaces, examined for the Duke (**b**), MIDA (**c**), and reduced Duke (**d**) models. Dashed lines denote the worst case of the optimization among the rotations.

## Discussion

Our metasurface design targets the homogenization of the magnetic field generated from the cylindrical transmit coil of the size 32 cm inner diameter and 24 cm height. This footprint is larger than those of conventional coils, such as the FDA-approved 1Tx/32Rx 7 Tesla head coil[35] of 30.4 cm inner diameter and 20 cm height, to cover both the entire head and the newly introduced metasurface objects. Because the magnetic field ansatz depends on the coil design, the metasurface structure should also be modified when we employ conventional coils (Supplementary Note S15).

To realize more compact UHF MRI with smaller coils, the meta-atoms that comprise metasurfaces need to be replaced to achieve wider angular coverage considering the changed solid angle of a human head with respect to each meta-atom. On the other side of MRI developments, considering recent advancements in portable low field (LF) MRI with $B_0 \leq 0.5T$[36,37], our results providing designed $\mathbf{B}_1^+$ field distribution could enhance both noise cancellation and image reconstruction essential in LF MRI.

Our results also inspire a novel perspective on the human brain as correlated disorder for wave physics. Although the structural and functional network correlations of the brain have been intensively examined with respect to MRI treatments[38], its material correlation in the realm of wave physics has yet to be discussed. The nonuniform profile of metasurface objects after the evolutionary pruning optimization demonstrates that the brain is statistically nonuniform in terms of its permittivity and conductivity. Furthermore, the stable operation of our optimized design for different head models implies that the hierarchical and correlated network of a human brain is reflected in terms of the shared correlation for electromagnetic waves.

In terms of optical design, our results introduce a new class of constant-intensity waves within non-Hermitian systems. In contrast to previous studies in this field[13-15], the field homogenization in the MRI exhibits critical differences, including noninvasive ROI filled with uncontrollable multiphase materials. We demonstrate that our semi-analytical and numerical approach is efficient for this intricate design problem. The extension of our method to other critical applications in biophotonics and medical treatment, such as the concentration of wave energy to a specific region of biological tissues, is straightforward.

In conclusion, we developed a semi-analytical method based on scattering theory to numerically design metasurface structures for volumetric $\mathbf{B}_1^+$ homogenization in the UHF brain

MRI. The core of this method involves characterizing the magnetic field ansatz to describe scattering from each metasurface object. This ansatz enables systematic optimization through the combined use of gradient descent and pruning techniques, leading to high-performance volumetric homogenization across different head models. Notably, the optimized platform operates in off-resonant states with enhanced magnetic energy, which contributes to further improvements in the safety issue by decreasing the Joule heating. Our systematic design process offers a general recipe for implementing stable and safe UHF MRI systems beyond 7T, while also allowing for noninvasive field flattening or concentration essential for medical applications of electromagnetic and acoustic waves. The result also inspires the field of disordered photonics[39] in terms of scattering profile manipulations using the field ansatz and evolutionary pruning.

## Data availability

The data that support the plots and other findings of this study are available from the corresponding author upon request.

## Code availability

All code developed in this work will be made available upon request.

## Acknowledgements

We acknowledge financial support from the National Research Foundation of Korea (NRF) through the Mid-career Researcher Program (No. RS-2023-00274348) and Global Frontier Program (No. 2014M3A6B3063708), funded by the Korean government.

## Author Contributions

N.P. conceived the idea for field homogenization and heat reduction in the UHF MRI. G.Y., S.Y., and N.P. developed the analytical framework. G.Y. developed the numerical tool and conducted the optimization under the supervision of N.P. J.L. defined the necessary performance criteria for UHF MRI. G.Y., S.Y., and N.P. analysed the data. N.P. oversaw the project. All authors discussed the results and contributed to the final manuscript.

## Competing interests

A provisional patent application (KR Prov. App. 10-2024-0028542) has been filed by Seoul National University. The inventors include G.Y., S.Y. and N.P. The application, which is pending, contains proposals of $\mathbf{B}_1^+$ field homogenization in UHF MRI.

## Additional information

**Correspondence and requests for materials** should be addressed to S.Y. or N.P.

# Supplementary Information for "Volumetric $B_1^+$ field homogenization in 7 Tesla brain MRI using metasurface scattering"

Gyoungsub Yoon[1], Sunkyu Yu[2,†], Jongho Lee[3], and Namkyoo Park[1,*]

[1]Photonic Systems Laboratory, Department of Electrical and Computer Engineering, Seoul National University, Seoul 08826, Korea

[2]Intelligent Wave Systems Laboratory, Department of Electrical and Computer Engineering, Seoul National University, Seoul 08826, Korea

[3]Laboratory for Imaging Science and Technology, Department of Electrical and Computer Engineering, Seoul National University, Seoul, Korea

E-mail address for correspondence: [†]sunkyu.yu@snu.ac.kr, [*]nkpark@snu.ac.kr

**Note S1. Transverse electromagnetic resonator**

**Note S2. Impedance matching circuit**

**Note S3. $B_1^+$ generation**

**Note S4. Lippmann-Schwinger equation for MRI environments**

**Note S5. Complex Poynting vector at the metasurface**

**Note S6. Theoretical planar metasurfaces**

**Note S7. Practical realization of metasurfaces**

**Note S8. Implementation of curved metasurfaces**

**Note S9. Target brain tissues**

**Note S10. Impact of pruning techniques**

**Note S11. Model-specified design of metasurfaces**

**Note S12. Field homogenization in MIDA and reduced Duke**

**Note S13. Robustness analysis**

**Note S14. Electromagnetic energy density inside heads**

**Note S15. Variations in coil designs**

**Note S1. Transverse electromagnetic resonator**

As a component of 7T brain MRI coils, a finite transverse electromagnetic (TEM) resonator is employed to generate radiofrequency (RF) magnetic field $\mathbf{B}_1^+$ (Fig. S1a), which is essential for exciting in vivo nuclear spins during image processing. The resonator (length $l = 24$ cm) consists of a dielectric layer of Teflon (permittivity $\varepsilon_r = 2.08$ and height $h = 2$ cm), embedded in between a copper strip ($w_s = 1.4$ cm, $t = 2$ mm) and a ground copper plate (width $w_g = 6$ cm and thickness $t = 2$ mm). According to transmission line theory, the operation of the TEM resonator is described by two parameters: the wavenumber $\beta$ and the characteristic impedance $Z_0$ (Fig. S1b). Two identical capacitors are tuned at both ends of the resonator to modify the interference patterns inside the structure. When the capacitance reaches the parallel-resonance condition, a symmetric current distribution is derived, thereby producing a uniform magnetic field throughout the imaging volume (Fig. S1c). In our coil design, the capacitance is set to 3.16 pF.

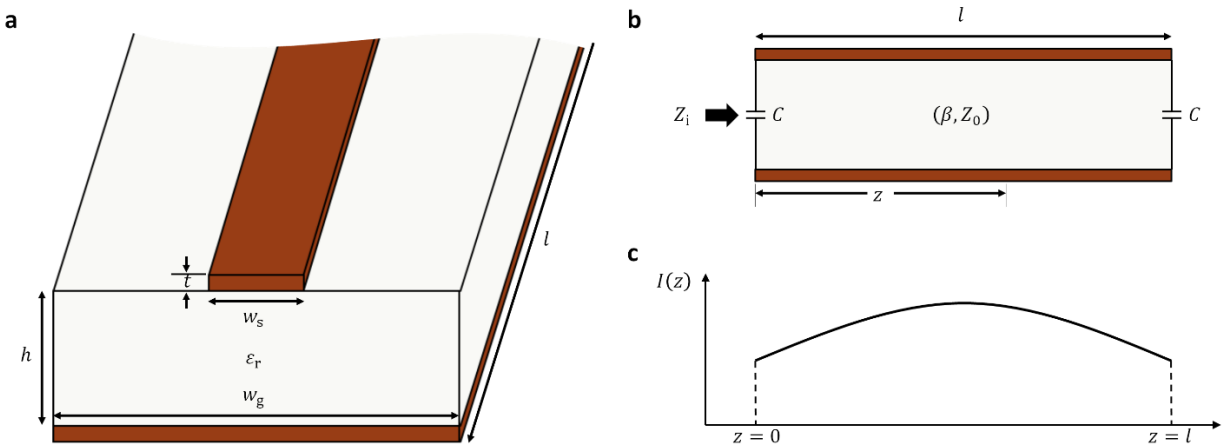

**Fig. S1. A TEM resonator. a,** Schematic of a TEM strip resonator. **b,** TEM strip resonator with wave properties: the wavenumber $\beta$ and the characteristic impedance $Z_0$. **c,** Current distribution along the capacitively shunted TEM strip.

**Note S2. Impedance matching circuit**

In our numerical analysis, a voltage source with an internal resistance $R_s$ of 50 Ω excites a TEM resonator (Fig. S2a). The wave reflection coefficient Γ at the source termination is given by

$$\Gamma = \frac{Z_p - R_s}{Z_p + R_s} \tag{S1}$$

where $Z_p$ (= $V_p/I_p$) represents the impedance toward the input port. For the maximum power transfer, $Z_p$ must be matched to $R_s$. For this, an impedance matching circuit, composed of one inductor and one capacitor, is placed between the source and the input port (Fig. S2b). The values of the lumped elements are finely tuned until the return loss at the source termination falls below –40 dB (Figs S2c and S2d). Figures S2e and S2f show the comparison of the numerically calculated magnetic field radiation from the TEM strip resonator without and with the impedance matching circuit.

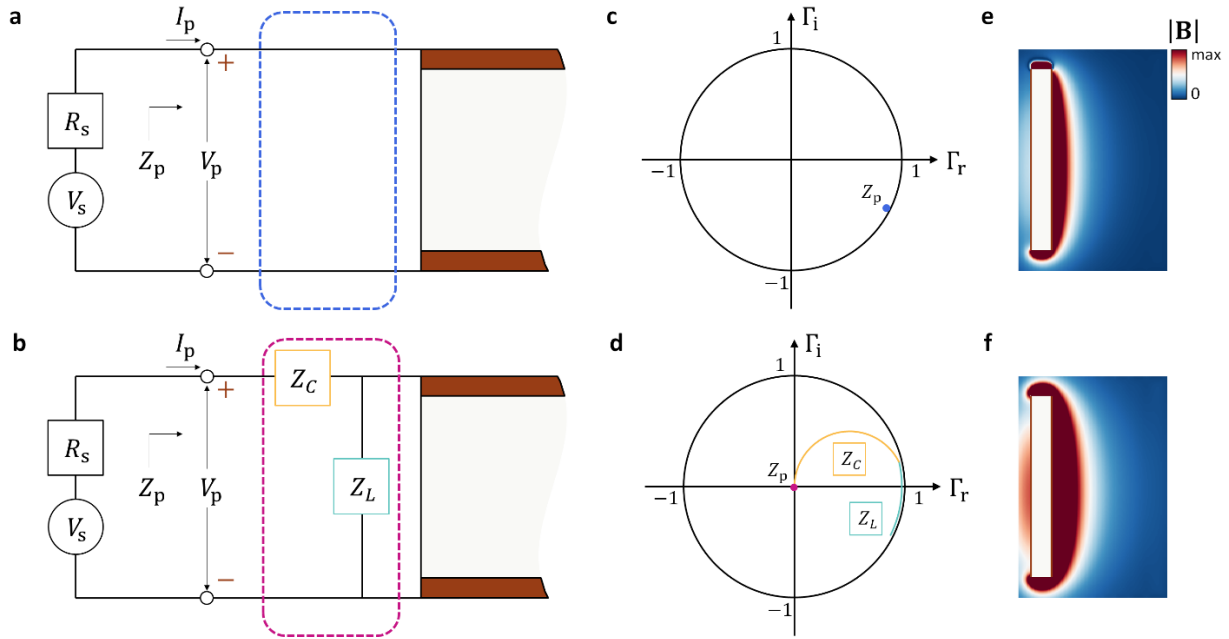

**Fig. S2. Impedance matching circuit for the maximum power transfer.** TEM strip resonator (**a,b**), Smith charts (**c,d**), and magnetic field radiations (**e,f**) without (**a,c,e**) and with (**b,d,f**) the matching circuit.

## Note S3. $B_1^+$ generation

In our coil design, the 16 identical TEM resonators are evenly spaced along the azimuthal axis and are excited by sequentially phase shifted 16 voltage sources. This configuration collectively produces circularly polarized magnetic fields (Fig. 3a), as follows:

$$B_1^+ = \frac{B_x + jB_y}{2}. \tag{S2}$$

We use the vector notation $\mathbf{B}_1^+$ for the RF magnetic field to reflect its vectorial nature. The real and imaginary parts of the vector field correspond to the $x$ and $y$ components of the magnetic field projected onto a coordinate frame rotating at an angular Larmor frequency $\omega$ (Fig. S3a). Uniform $\mathbf{B}_1^+$ distribution within the coil is shown in Figs. S3b and S3c. The coil transfers the majority of energy as the circular polarization, which is demonstrated by comparing $|\mathbf{B}_1^+|$ (DC value) with $B_{rms}$ within the coil (Figs. S3b and S3d).

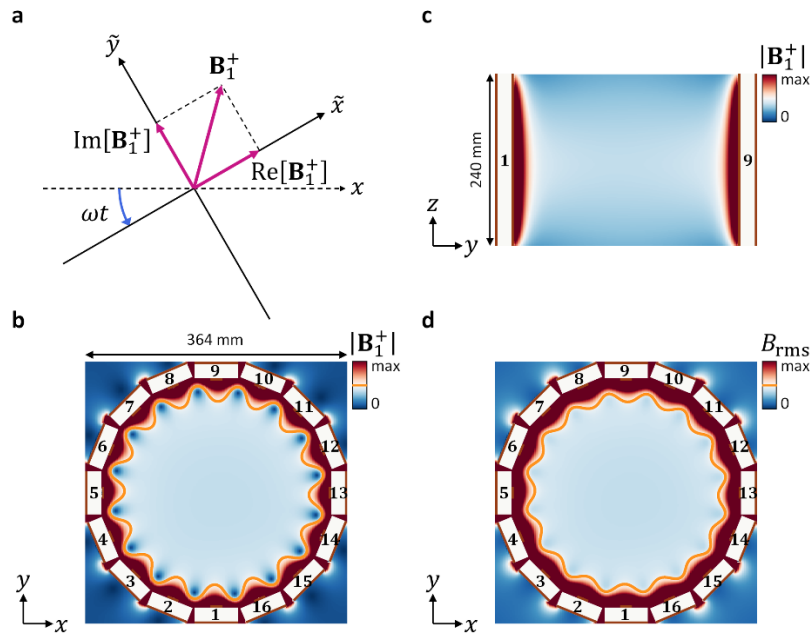

**Fig. S3. 16-channel TEM coil. a,** $\mathbf{B}_1^+$ onto a coordinate frame rotating at an angular Larmor frequency $\omega$. **b,c,** Distribution of $\mathbf{B}_1^+$ generated by a 16-channel TEM coil in the $x$-$y$ (**b**) and $y$-$z$ (**c**) planes. The phase of the $n$th source is given by $2\pi n/16$. **d,** Distribution of $B_{rms}$ within the coil

in the *x-y* plane. The orange lines indicate contours of equal magnitude in **b** and **d**.

## Note S4. Lippmann-Schwinger equation for MRI environments

We assume 7T MRI environments with the Maxwell's equations under linear and isotropic media with the source-free condition:

$$\nabla \times \mathbf{E}(\mathbf{r}) = -j\omega \mathbf{B}(\mathbf{r}) \qquad \nabla \cdot (\varepsilon_0 \varepsilon_r(\mathbf{r})\mathbf{E}(\mathbf{r})) = 0 \tag{S3}$$

$$\nabla \times \mathbf{B}(\mathbf{r}) = j\omega\mu_0\varepsilon_0\varepsilon_r(\mathbf{r})\mathbf{E}(\mathbf{r}) \qquad \nabla \cdot \mathbf{B}(\mathbf{r}) = 0.$$

In Eq. (S3), $\mathbf{E}$ and $\mathbf{B}$ are the electric field and magnetic flux density, $\varepsilon_0$ and $\mu_0$ are the vacuum permittivity and permeability, and $\varepsilon_r$ is the relative permittivity at the angular Larmor frequency $\omega$. Equation (S3) leads to the following second-order wave equation:

$$\nabla^2 \mathbf{B}(\mathbf{r}) + k_0^2 \varepsilon_r(\mathbf{r})\mathbf{B}(\mathbf{r}) + (\nabla \varepsilon_r(\mathbf{r})) \times \frac{\nabla \times \mathbf{B}(\mathbf{r})}{\varepsilon_r(\mathbf{r})} = 0 \tag{S4}$$

where $k_0 = \omega\sqrt{\mu_0\varepsilon_0}$. When we assume $\nabla \varepsilon_r(\mathbf{r})$ is negligible in our scenarios, such as the MRI with a phantom with slowly-varying optical parameters, the governing equation becomes the homogeneous vector Helmholtz equation:

$$(\nabla^2 + k_0^2 \varepsilon_r(\mathbf{r}))\mathbf{B}(\mathbf{r}) \sim 0. \tag{S5}$$

In the presence of scatterers, which are the metasurface objects in our work, $\mathbf{B}$ and $\varepsilon_r$ can be decomposed into the superposition of incident and scattering configurations:

$$\mathbf{B}(\mathbf{r}) = \mathbf{B}_0(\mathbf{r}) + \mathbf{B}_s(\mathbf{r}) \qquad \varepsilon_r(\mathbf{r}) = \varepsilon_{r0} + \varepsilon_M(\mathbf{r}). \tag{S6}$$

Here, $\mathbf{B}_0$ is the magnetic field generated by a coil, $\mathbf{B}_s$ is the magnetic field scattered from the metasurface, $\mathbf{B}$ is the total magnetic field, $\varepsilon_{r0}$ is the relative permittivity of the background medium, and $\varepsilon_M$ is the permittivity perturbation due to the metasurface objects. By substituting Eq. (S6) into Eq. (S5), we obtain

$$(\nabla^2 + k_0^2 \varepsilon_{r0}(\mathbf{r}))\mathbf{B}_s(\mathbf{r}) \sim -k_0^2 \varepsilon_M(\mathbf{r})\mathbf{B}(\mathbf{r}) \tag{S7}$$

The solution to Eq. (S7) can be formulated using the Green's function $\mathbf{G}$, which satisfies the

following equation:

$$(\nabla^2 + k_0^2 \varepsilon_{r0}(\mathbf{r}-\mathbf{r}'))\mathbf{G}(\mathbf{r}) \sim -4\pi\delta(\mathbf{r}-\mathbf{r}')(\hat{\mathbf{x}}+\hat{\mathbf{y}}+\hat{\mathbf{z}}) \tag{S8}$$

where $\delta$ denotes the Dirac delta function. Equation (S8) can be reformulated into the equivalent Lippmann-Schwinger equation shown in Eq. (1) in the main text.

**Note S5. Complex Poynting vector at the metasurface**

Due to the limited space available within an MRI coil, metasurfaces are positioned in the subspace $\Omega$, adjacent to the coil (Fig. S4a). To identify the major power component that drives the positioned metasurfaces, we analyse the complex Poynting vector

$$\mathbf{S} = \frac{\mathbf{E} \times \mathbf{H}^*}{2} \tag{S9}$$

where $\mathbf{E}$ is the complex electric field and $\mathbf{H}^*$ is the conjugate of the complex magnetic field. The driving field in $\Omega$ is mostly coupled to the waves oscillating along the TEM resonator (Fig. S1b) due to near-field interactions. Therefore, the imaginary reactive part of the Poynting vector is dominant due to the $\pi/2$ phase difference in the resonance condition of the TEM resonator (Figs. S4a and S4b). Moreover, the majority of the driving field oscillates along the z-axis which is perpendicular to the direction toward human heads (Fig. S4c).

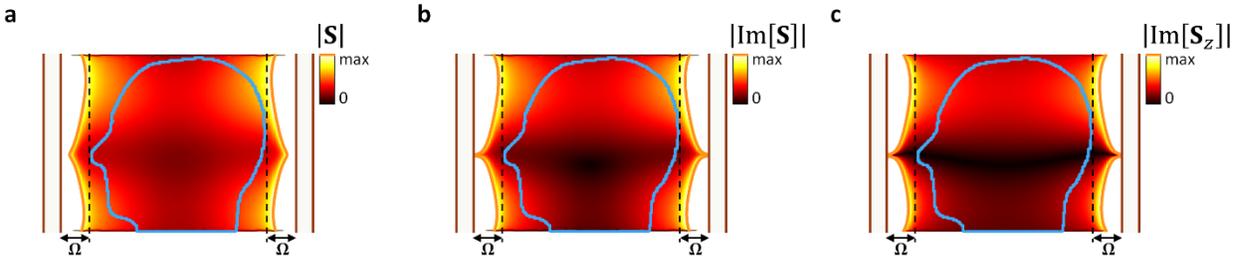

**Fig. S4. Complex Poynting vector within a coil. a-c,** Comparison of |**S**| (**a**), |Im[**S**]| (**b**), and |Im[**S**$_z$]| (**c**) with a virtual outline of a human head. **S**$_z$ is the z-axis component of the complex Poynting vector **S**. The orange lines indicate contours of the equal magnitude in **a-c**.

**Note S6. Theoretical design of metasurfaces**

In this note, we derive an analytical equation that enables the flexible design of a metasurface with high effective permittivity. To simplify the derivation, we employ two assumptions. First, the metasurface is assumed to be an infinite planar structure. Second, the in-plane wave propagates only along the $z$-axis. The corresponding wavenumber is then calculated to determine effective material properties.

The applied metasurface consists of periodically arranged cylindrical copper wires and parallel-plate capacitors with a unit length $\Lambda$ of 1 cm (Figs. S5a and S5b). The propagating wave induces currents in the structure (Fig. S5c), and the structure is simplified into an equivalent circuit where all inductances are mutually coupled to each other (Fig. S5d). By applying Kirchhoff's voltage law to one loop with mesh current $I_z$ in Fig. S5c, we derive the coupled equation

$$2\frac{I_z}{C} - (I_{z+1} + I_{z-1})\frac{1}{C} - \omega^2 \sum_n I_{z+n} M_{z+n} \Lambda = 0 \tag{S10}$$

where $C$ is capacitance, $n$ takes all integer values, and $M_{z+n}$ is the mutual inductance per unit horizontal length between the investigated loop and an infinite parallel-wire transmission line carrying current $I_{z+n}$. The mutual inductance $M_{z+n}$ is defined as

$$M_{z+n} = \begin{cases} \mu_0 \ln(\frac{\Lambda - r}{r}) & \text{when } n = 0 \\ \frac{\mu_0}{2\pi}\left( \ln\left(\frac{(n+1)\Lambda - r}{n\Lambda + r}\right) + \ln\left(\frac{(n-1)\Lambda + r}{n\Lambda - r}\right) \right) & \text{when } n \neq 0 \end{cases} \tag{S11}$$

where $r$ is the radius of the metallic wires. By applying Bloch theorem, we determine the relation between the neighbouring mesh currents and the wavenumber $\beta$, given by

$$I_{z+n} = I_z e^{-j\beta n \Lambda} \tag{S12}$$

Substitution of Eq. (S12) into Eq. (S10) yields the dispersion relation of the structure:

$$\omega = \frac{2}{\sqrt{CM'}} \sin\left(\frac{\beta\Lambda}{2}\right), \quad \text{where } M' = \sum_n e^{-j\beta n\Lambda} M_{z+n}\Lambda. \tag{S13}$$

To verify the equation, we compute the eigenmode and eigenvalue $\beta$ of the structure using a finite element method (FEM) software, COMSOL Multiphysics. As an example, when the capacitance $C$ becomes 45 pF with a relative permittivity of 3236 in Fig. S5b, we observe excellent agreement between the theoretical predictions and the numerical solutions within the half Brillouin zone ($0 < \beta\Lambda/2\pi < 0.5$) in Fig. S5e. Moreover, when the frequency is fixed at 300 MHz, operating frequency in 7T MRI, the wavenumber $\beta$ can be tuned by adjusting the capacitance $C$. To figure out the effective permittivity of the metasurface, we then compare its wavenumber $\beta$ with that of a dielectric slab waveguide of equal 2 mm thickness in the $y$-direction. The dispersion relation for transverse electric (TE) modes in the dielectric slab waveguide is obtained as follows:

$$\tan\left(\frac{h\sqrt{k_0^2 \varepsilon_r - \beta^2}}{2}\right) = \sqrt{\frac{\beta^2 - k_0^2}{k_0^2 \varepsilon_r - \beta^2}} \tag{S14}$$

where $h$ is the thickness of the dielectric slab, $\varepsilon_r$ is the relative permittivity, and $k_0 = \omega\sqrt{\mu_0 \varepsilon_0}$. Consequently, the high effective permittivity of the designed metasurface is validated by comparing the in-plane wavenumbers of the two structures (Fig. S5f). Our theoretical analysis explicitly reveals the effective material properties of the metasurface and demonstrates its potential for substituting all HPMs in UHF MRI applications.

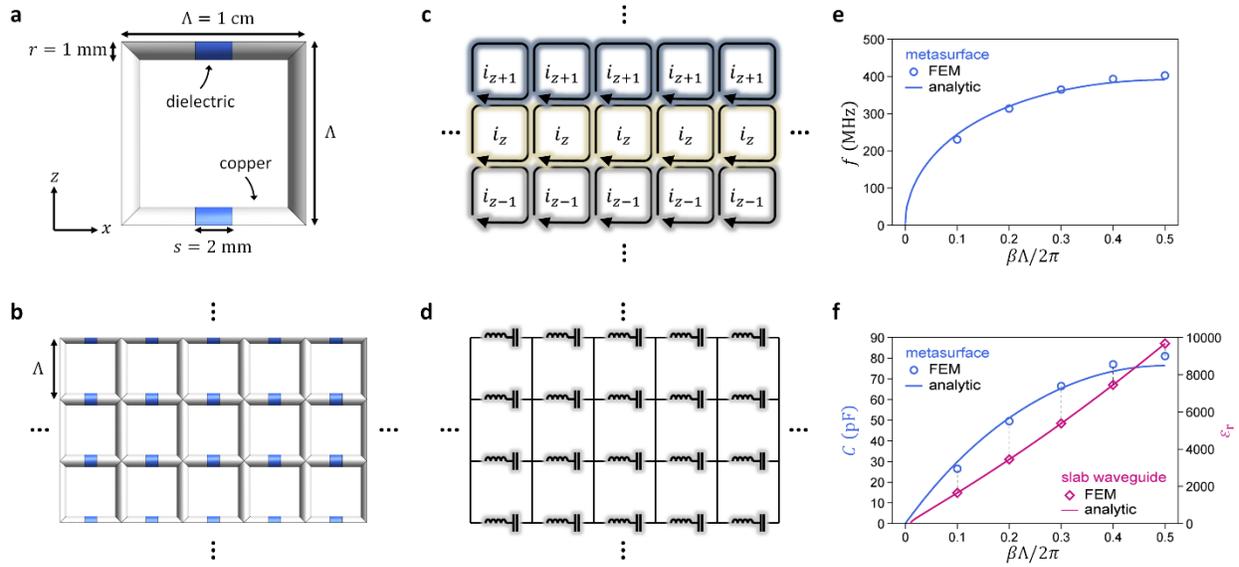

**Fig. S5. Effective material properties of the copper-wire metasurface. a-d,** Schematics of the metasurface structure: unit cell (**a**), two-dimensional array (**b**), current distribution (**c**), and equivalent circuit (**d**). **e-f,** Validation of the derived dispersion relation. Comparison of the theoretical predictions (lines) and the numerical solutions (dost) for a fixed capacitance (**e**) and a fixed frequency (**f**).

## Note S7. Practical realization of metasurfaces

As mentioned in Supplementary Note S6, achieving a 45 pF capacitance requires a dielectric material with a relatively high permittivity of 3236 (Fig. S5b). In this note, we simplify the structure to enable its practical realization with acceptable material properties.

The invariant current phase along the transverse direction in Fig. S5c disrupts the current paths through the vertical wires, effectively creating open circuits between each horizontal wire (Fig. S6a). This configuration allows for the series connection of $m$ capacitors along the wire, which reduces the relative permittivity by a factor of $m$ (Fig. S6b). We then utilize the extended space by increasing the surface area of the capacitor electrodes (Fig. S6c). Rectangular wires replace the cylindrical ones with the radius ($r$), covering the dielectric material with the width ($w$), height ($s$), and thickness ($t$). This metasurface configuration reduces the relative permittivity by a factor of $n$:

$$n = \frac{wt}{\pi r^2}, \tag{S15}$$

which corresponds to the ratio of the surface areas. Although adjusting the electrode separation ($s$) could further reduce the relative permittivity, we fix the gap at 2 mm to align with the 2 mm$^3$ grid resolution used in the MRI simulation. As a result, the overall reduction factor of the necessary becomes $mn$.

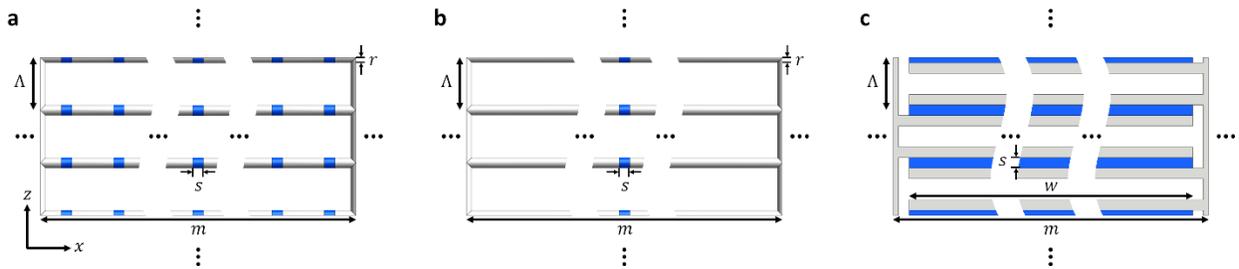

**Fig. S6. Simplified metasurfaces for practical implementation. a,** Initial metasurface structure. **b,** Series connection of $m$ capacitors. **c,** Extension of the surface area of capacitor plates.

**Note S8. Implementation of curved metasurfaces**

Figure S7 shows cylindrical metasurfaces designed through the process in previous notes. The metasurfaces share a common diameter (26 cm), but vary in length to accommodate different half-wavelength resonances. Due to the discrete rotational symmetry of RF fields (Fig. S3b), *m* and *w* (Fig. S6c) become 5.1 cm and 4.4 cm, respectively. The required relative permittivity is calculated to be 23 for the 3 cm half-wavelength of in-plane waves. The substructure, consisting of three unit cells along the *z*-axis, is employed for the 3 cm half-wavelength resonator (Fig. S7a). Due to the slight deviation between theoretical predictions and numerical solutions (Fig. S5f), we search for the exact relative permittivity near 23, which leads to resonance peak in the structure. As a result, the value of 25 achieves this. In the similar manner, we design another metasurface for the 2 cm half-wavelength resonator (Fig. S7b). Their penetration depths, which generally depend on the effective wavelengths in the structures, are then compared (Figs. S7c and S7d). We conclude that the 3 cm half-wavelength resonator is more suitable for exciting human heads which will be positioned within the structure. Moreover, the metasurfaces demonstrate three half-wavelength resonances when their lengths are adjusted to include three times the number of unit cells (Figs. S7e and S7f). This design further validates our theory-based design process.

For efficient field coupling with ellipsoidal human heads, three elliptical metasurfaces are designed for Duke, MIDA, and reduced Duke, respectively (Fig. S7g). Supplementary Table 1 provides the design parameters for each structure.

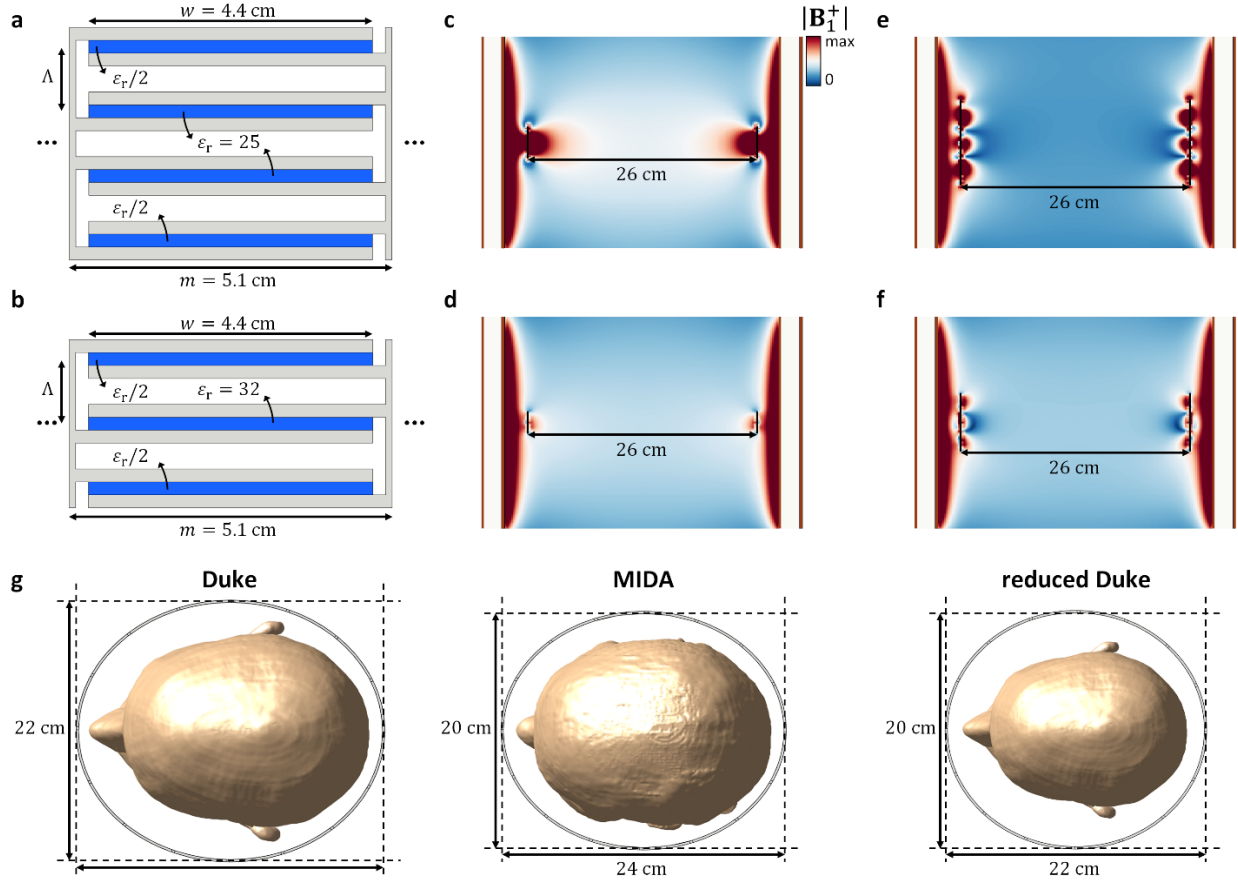

**Fig. S7. Implementation of curved metasurfaces. a,b,** Design of cylindrical metasurfaces (26 cm diameter) for 3 cm (**a**) and 2 cm (**b**) half-wavelength resonances. **c,d,** $B_1^+$ distribution with half-wavelength resonators. **e,f,** $B_1^+$ distribution with metasurfaces, where the lengths are adjusted to include three times the number of unit cells. The structural and material details are shown in **a** and **b**.

**Supplementary Table 1. Metasurface design parameters for different head models.**

| Models | Parameters | | |
|---|---|---|---|
| | $\varepsilon_r$ | $w$ [cm] | $m$ [cm] |
| Duke | 32 | 4.0 | 4.7 |
| MIDA | 40 | 3.6 | 4.3 |
| Reduced Duke | 46 | 3.4 | 4.1 |

## Note S9. Target brain tissues

Supplementary Table 2 lists the target brain tissues for the field homogenization and their material properties—relative permittivity $\varepsilon_r$ and electric conductivity $\sigma$. The exact name of each tissue could vary depending on both gender and adapted versions of anatomical models.

**Supplementary Table 2. Target brain tissues with their material properties.**

| Tissues | $\varepsilon_r$ | $\sigma$ [S/m] |
|---|---|---|
| artery | 65.7 | 1.3 |
| cerebellum | 59.7 | 1.0 |
| cerebrospinal fluid | 72.7 | 2.2 |
| cerebrum grey matter | 60.0 | 0.7 |
| cerebrum white matter | 43.8 | 0.4 |
| commissura | 43.8 | 0.4 |
| corpus callosum | 43.8 | 0.4 |
| dura mater | 48.0 | 0.8 |
| hippocampus | 60.0 | 0.7 |
| hypophysis | 62.4 | 0.9 |
| hypothalamus | 60.0 | 0.7 |
| medulla oblongata | 59.7 | 1.0 |
| midbrain | 59.7 | 1.0 |
| nerve | 36.9 | 0.4 |
| pineal body | 62.4 | 0.9 |
| pons | 59.7 | 1.0 |
| thalamus | 60.0 | 0.7 |
| vein | 65.7 | 1.3 |

**Note S10. Impact of pruning techniques**

Pruning of the $N$ metasurface objects during the optimization process (Fig. 2b in the main text) leads to a simpler design of the metasurface and mitigates overfitting to a particular target domain for the field homogenization. Supplementary Table 3 presents the CV of Duke, which is computed by applying the optimal weights $\{c_n\}_{Duke}$ to the basis $\{\mathbf{B}_n\}_{Duke}$. It also shows the CV of MIDA and reduced Duke by applying the $\{c_n\}_{Duke}$ to their bases, $\{\mathbf{B}_n\}_{MIDA}$ and $\{\mathbf{B}_n\}_{reduced\ Duke}$. We note that all bases $\{\mathbf{B}_n\}$ are obtained using the same metasurface object tailored for Duke (Fig. S7g) to investigate the sensitivity of the designed field patterns under variations in head models. To quantify the sensitivity, we calculate mean absolute deviation (MAD), $\text{mean}(|X_i - \bar{X}|)$, of the CV values for a given $M$ number of the metasurface objects—a low MAD indicates high compatibility of the corresponding metasurface arrangement. In the case of Duke, increase in the number of basis vectors results in decrease in CV, whereas this tendency is not observed in the other models due to overfitting. Therefore, the use of four metasurface objects offers the lowest sensitivity against variations in head models, while maintaining excellent field homogeneity.

**Supplementary Table 3. CVs and MADs for a given $M$ number of the metasurface objects.**

| $M$ | Duke | MIDA | Reduced Duke | MAD [$10^{-1}$] |
|---|---|---|---|---|
| 6 | 0.136 | 0.196 | 0.228 | 0.337 |
| 5 | 0.137 | 0.182 | 0.218 | 0.279 |
| 4 | 0.139 | 0.186 | 0.216 | 0.275 |
| 3 | 0.166 | 0.212 | 0.294 | 0.476 |

## Note S11. Model-specified design of metasurfaces

The exact length of the metasurface object in Fig. S7a is 3.6 cm. Therefore, structural overlap arises when the metasurface object excites human heads at different positions spaced every 3 cm (Fig. 2a in the main text). Figure S8 shows the realization of a metasurface consisting of two adjacent objects, indexed as $m$ (1,2). Notably, the relative permittivity in the overlapping region becomes $(\varepsilon_{r1} + \varepsilon_{r2})/2$, which demonstrates the equivalent of two edge capacitors connected in parallel (Fig. S8b). Supplementary Table 4 presents the relative permittivities of the remained four metasurface objects tailored to Duke, MIDA, and reduced Duke (Fig. S7g) with reordered indices (Fig. 2c in the main text). It is noteworthy that the material properties can be further controlled by adjusting the separation gap between the electrodes covering dielectric materials.

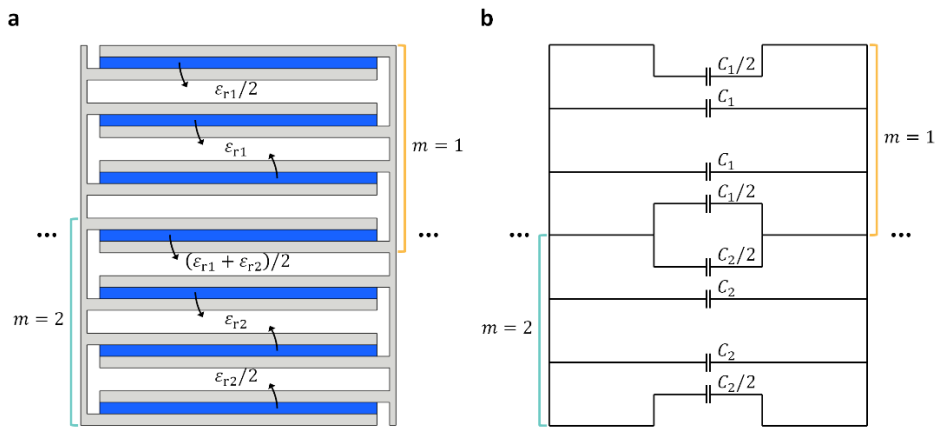

**Fig. S8. Integration of metasurface objects. a,b,** Integration of two adjacent metasurface objects with illustration of the assigned material properties (**a**), and the equivalent circuit (**b**)

**Supplementary Table 4. Relative permittivities of the $m$th metasurface object.**

| $m$ | Duke | MIDA | Reduced Duke |
|---|---|---|---|
| 1 | 23.1 | 20.0 | 18.4 |
| 2 | 5.5 | 8.0 | 9.2 |
| 3 | 22.0 | 32.0 | 36.8 |
| 4 | 12.1 | 12.0 | 13.8 |

## Note S12. Field homogenization in MIDA and reduced Duke

Figure S9 shows the field homogenization for MIDA (Figs. S9a-c) and reduced Duke (Figs. S9d-f), both with and without the use of metasurfaces tailored to each head. Their sizes and design parameters are detailed in Supplementary Notes S8 and S11.

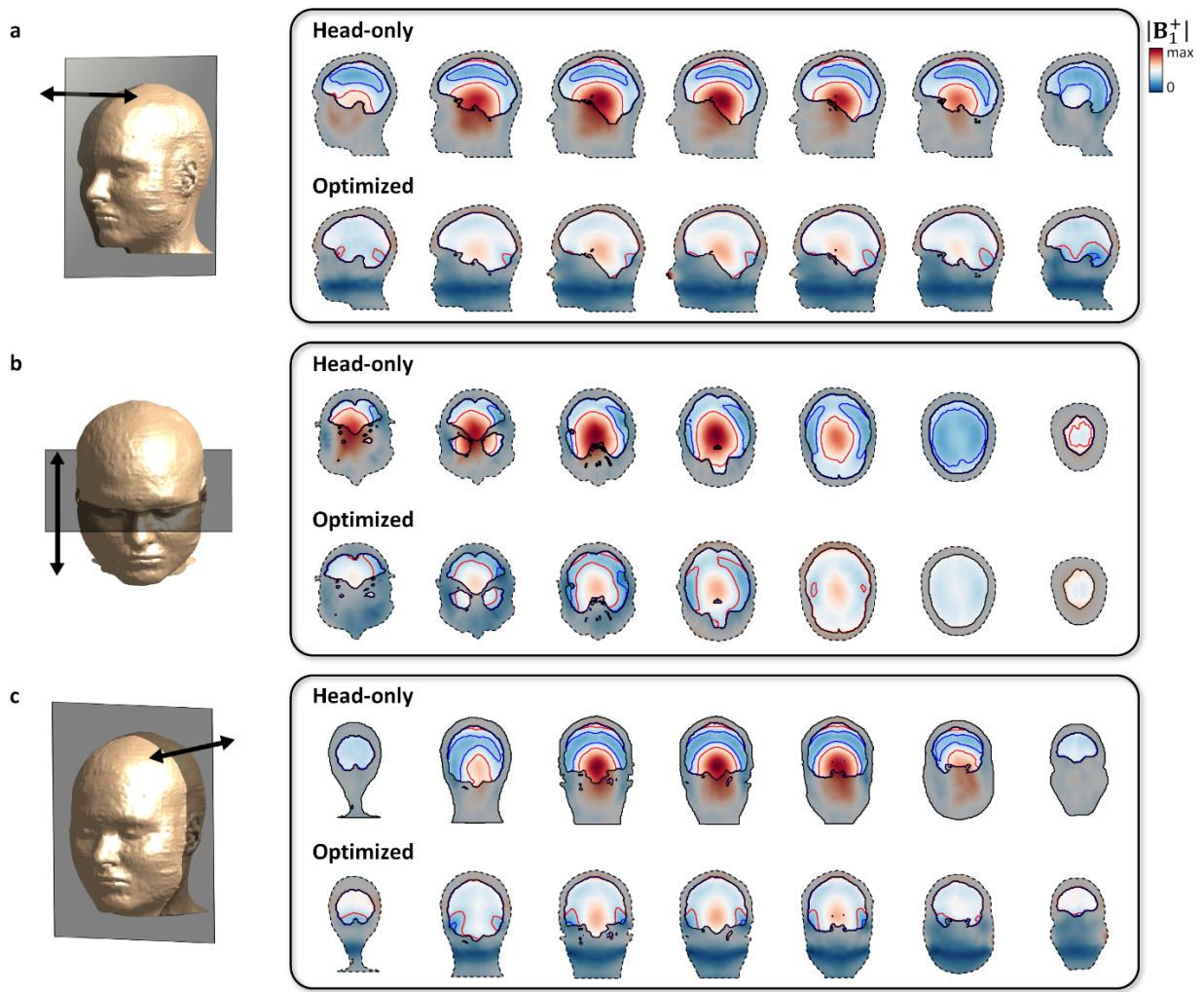

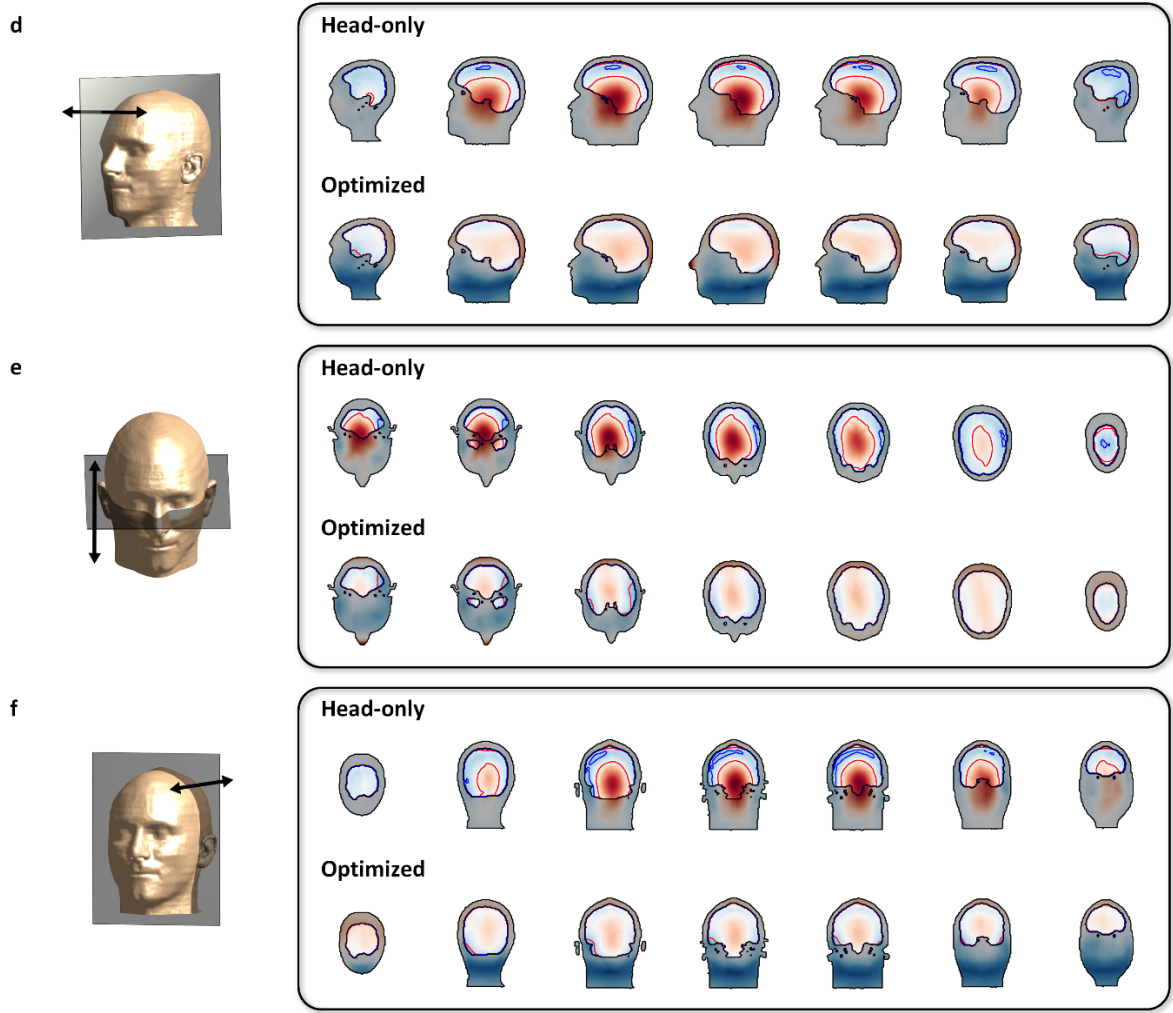

**Fig. S9. UHF MRI performance on field homogenization. a-f**, Comparisons of the $B_1^+$ distributions in MIDA (**a-c**) and reduced Duke (**d-f**) models without (upper) and with (lower) metasurface objects: Representative $B_1^+$ slices at different para-sagittal planes (**a,d**), para-axial planes (**b,e**), and para-coronal planes (**c,f**). The $B_1^+$ distributions of all cases are normalized to the volume integral of $|B_1^+|$. The red and blue contours highlight the $|B_1^+| / |B_1^+|_{max}$ = 1/2 and 1/3, respectively.

**Note S13. Robustness analysis**

Figure S10 shows the statistical analysis of the $|\mathbf{B}_1^+|$ field in the ROI and the SAR in the entire head of MIDA and reduced Duke when the metasurface optimized for Duke is applied to them. The metasurface design demonstrates high performance for CV (Figs. S10a and S10b) and SAR (Figs. S10c and S10d) despite the variations in head models.

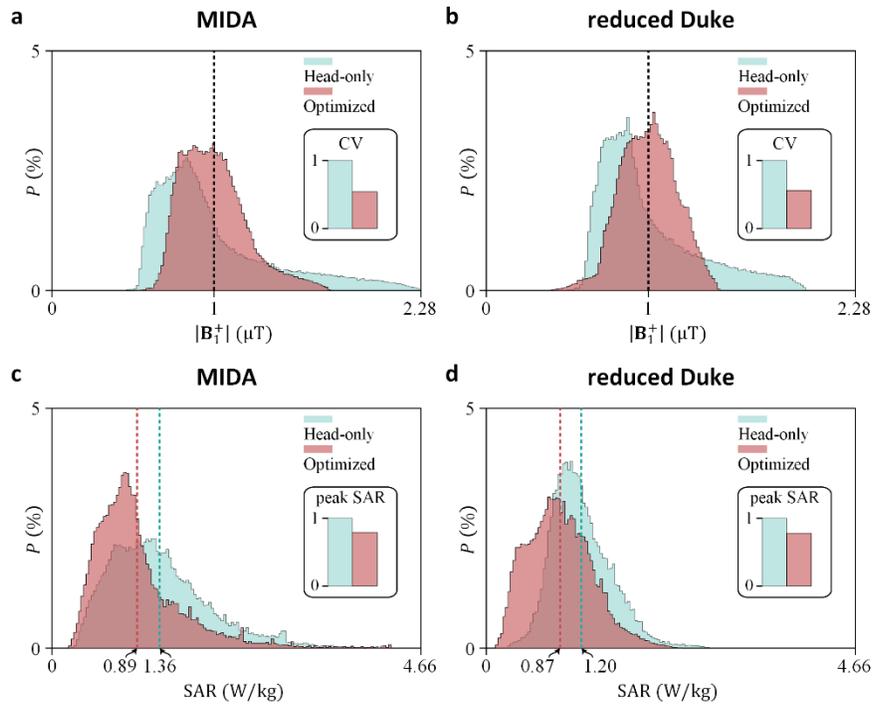

**Fig. S10. Robust performance for CV and SAR. a,b,** The volumetric probabilistic functions for the $|\mathbf{B}_1^+|$ with the average normalized to 1 μT, for the MIDA (**a**) and reduced Duke (**b**) models. **c,d,** The volumetric probabilistic functions for the SAR across the entire head of the MIDA (**c**) and reduced Duke (**d**) models. The improvement in the CV (**a-b**) and peak SAR (**c-d**) achieved with the optimized metasurfaces are shown on each inset, where the head-only model is normalized. The dashed lines in **a-d** denote the measured averages.

**Note S14. Electromagnetic energy density inside heads**

Figure S11 shows the volumetric probability functions for the electromagnetic energy density of Duke, MIDA, and reduced Duke. The energy density for complex electromagnetic fields is

$$U = \frac{1}{4}\left[\mathbf{E}(\mathbf{r})\cdot\mathbf{D}^*(\mathbf{r}) + \mathbf{H}(\mathbf{r})\cdot\mathbf{B}^*(\mathbf{r})\right] \quad (\mathrm{J/m^3}) \tag{S16}$$

where the first term is electric energy density $U_E$, and the second term is magnetic energy density $U_B$. The total energy density in the entire head, required to attain an average $|\mathbf{B}_1^+|$ of 1 µT in the ROI, decreases (Fig. S11a), while the ratio of $U_B$ to $U_E$ simultaneously increases when applying metasurfaces (Optimized cases) compared to head-only cases (Fig. S11b). The decreased total amount of $U_E$ contributes to the improvement of SAR-related performance.

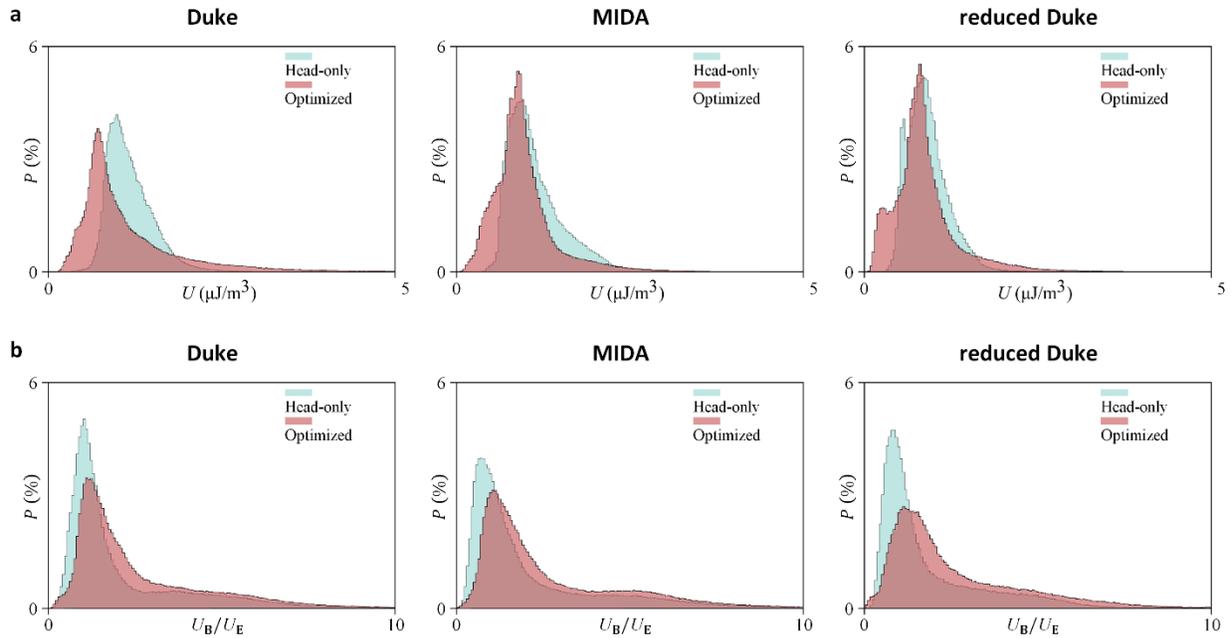

**Fig. S11. Analysis of electromagnetic energy density. a,** Volumetric probabilistic functions for the electromagnetic energy density $U$ in the entire head of Duke, MIDA, and reduced Duke models. **b,** Volumetric probabilistic functions for the ratio of $U_B$ to $U_E$ of Duke, MIDA, and reduced Duke models.

## Note S15. Variations in coil designs

Figure S12 shows whether variations in coil designs preserve the field homogeneity achieved with our 16-channel TEM coil and the optimized metasurface for Duke (left in Figs 12a-c). First, we apply another 16-channel TEM coil with mechanical dimensions different from our original coil design (middle in Figs 12a-c). Second, we apply a different type of coil, the high-pass birdcage coil (right in Figs 12a-c). Two additional coils have a common inner diameter of 304 mm and a length of 20 mm, matching the birdcage coil component of the commercially available 7T head coil produced by Nova Medical. The results show that the field patterns obtained with the identical metasurface are sensitivity to coil designs. The variations in the incident field generated by coils distort the resulting scattering patterns inside both the metasurface and the ROI.

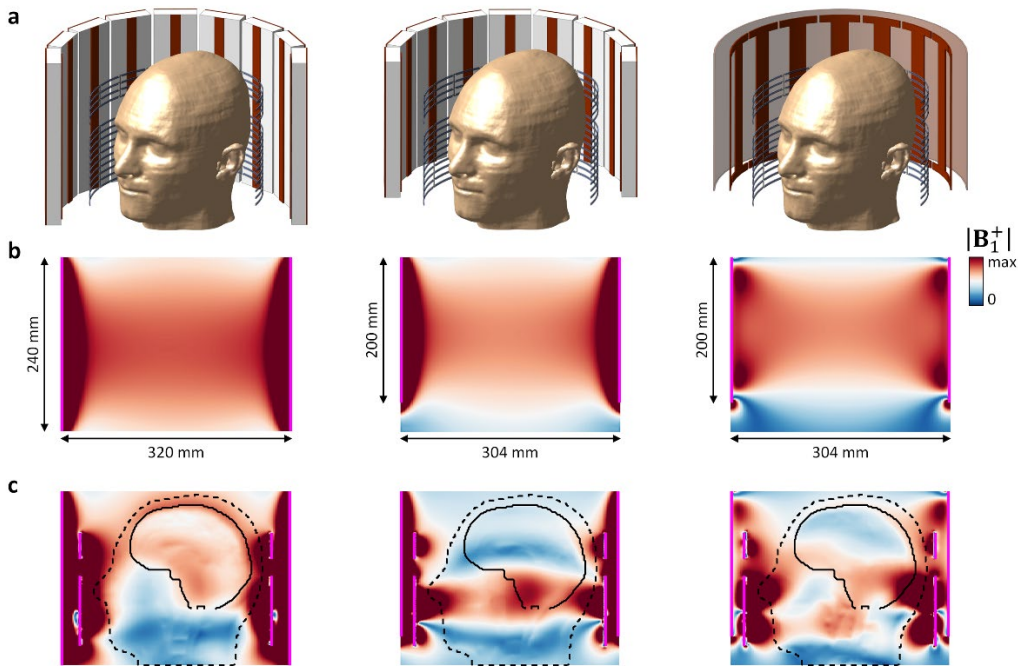

**Fig. S12. Comparison of coil desings. a,** Variations in coil designs. **b,** Distribution of $B_1^+$ generated by coils only. **c,** Distribution of $B_1^+$ with the identical metasurface and Duke. The original TEM coil (left), small TEM coil (middle), and high-pass birdcage coil (right) are compared in **a-c**. Magenta lines indicate the inner boundaries of each coil and meatsurface **b-c**. Black dashed and solid lines indicate the contours of Duke and the ROI, respectively.